# Forecasting Cryptocurrency Returns from Sentiment Signals: An Analysis of BERT Classifiers and Weak Supervision

Duygu Ider, Stefan Lessmann

School of Business and Economics, Humboldt University of Berlin, Unter-den-Linden 6, 10099 Berlin, Germany

E-mail addresses: duygu.ider@hotmail.com (Duygu Ider), stefan.lessmann@hu-berlin.de (Stefan Lessmann)

Corresponding Author: Duygu Ider, Unter-den-Linden 6, 10099 Berlin, Germany. Tel. +49 15257162290; E-mail duygu.ider@hotmail.com

**Abstract:** Anticipating price developments in financial markets is a topic of continued interest in forecasting. Funneled by advancements in deep learning and natural language processing (NLP) together with the availability of vast amounts of textual data in form of news articles, social media postings, etc., an increasing number of studies incorporate text-based predictors in forecasting models. We contribute to this literature by introducing weak learning, a recently proposed NLP approach to address the problem that text data is unlabeled. Without a dependent variable, it is not possible to finetune pretrained NLP models on a custom corpus. We confirm that finetuning using weak labels enhances the predictive value of text-based features and raises forecast accuracy in the context of predicting cryptocurrency returns. More fundamentally, the modeling paradigm we present, weak labeling domain-specific text and finetuning pretrained NLP models, is universally applicable in (financial) forecasting and unlocks new ways to leverage text data.

# 1. Introduction

Cryptocurrency markets are considered volatile, risky, and prone to overreaction (Stavroyiannis, 2018). The rising popularity of cryptocurrency trading is accompanied by a constant flow of commentary on expected price developments. Numerous studies examine the level of influence of news and social media when predicting future stock market movements (Bollen, Mao, & Zeng, 2011; Antweiler & Frank, 2004; Kearney & Liu, 2014). Building on research that focuses on stock markets, the objective of this paper is to examine whether and to what extent the sentiment from news and social media contributes to the predictability of future cryptocurrency prices using machine learning (ML) models. To that end, we consider textual data from scraped news articles, Reddit posts, and Tweets related to the two largest cryptocurrencies by market capitalization, Bitcoin and Ethereum.

To distill information from text data, we consider a sentiment analysis framework and estimate sentiment scores using the popular BERT, *Bidirectional Encoder Representations from Transformers* (Devlin, Chang, Lee, & Toutanova, 2018), classifier. We then incorporate the sentiment scores in a daily return prediction model to clarify the predictive information of the text data vis-a-vis other factors and test whether it improves trading decisions. To achieve this, we simulate trading directional model forecasts over multiple test periods, which cover bullish and bearish market patterns.

A distinctive contribution of our study concerns the use of weak supervision. Using sentiment classifiers for financial forecasting involves a major challenge. Relevant (i.e., financial) text data with known sentiment labels is scarce. Exactly such labeled financial text data is essential to instantiate the state-of-the-art approach in natural language processing (NLP), which involves starting from a pretrained model like BERT and finetuning that model's parameters using text data from the target domain. We address this challenge by introducing weak supervision to the field of financial forecasting. We test the potential of weak supervision to train BERT on unlabeled, cryptocurrency-related news and social media dataset by introducing pseudo-labels from a zero-shot classifier. We verify empirically that BERT's task-specific and contextual learning ability is not compromised by the weakness of the labels and confirm their value in a financial forecasting case study.

Further empirical contributions of our study are threefold. First, we combine sentiment scores from self-scraped news articles, Tweets, and Reddit posts to forecast cryptocurrency returns. To our best knowledge, this rich combination of text sources has not been considered in prior work on cryptocurrency return prediction. Second, we collectively evaluate the predictive performance of various financial ML models with and without the extracted sentiment features. Based on those models, we examine the potential profit gains from trading those models' predictions in the scope of a specific trading strategy. Lastly, we evaluate multiple trading periods within a test period to create a distribution of investment returns, which allows less time-dependent predicted returns and more reliable statistical comparisons between models.

## 2. Literature Review

A large body of literature examines the predictability of financial markets (Rapach & Zhou, 2013, pp. 328–383; Granger, 1992). Motivations for corresponding research include testing the informational efficiency of specific markets (Nordhaus, 1987), benchmarking novel forecasting methods (Yu, Wang, & Lai, 2008), or devising algorithmic trading strategies (Brownlees, Cipollini, & Gallo, 2011). Recent studies increasingly rely on ML or deep learning for financial forecasting (Sezer, Gudelek, & Ozbayoglu, 2020). In the area of cryptocurrency forecasting, those techniques are especially popular and have shown better results than alternative approaches (Lahmiri & Bekiros, 2019; McNally, Roche, & Caton, 2018; Chen, Xu, Jia, & Gao, 2021).

Traditional financial markets have been established over the last century with an abundance of structured data available for research or commercial use (Wang, Lu, & Zhao, 2019). For cryptocurrency markets, structured data is not as readily available and the need for alternative data sources plays a key role in predicting returns. NLP approaches to text data for cryptocurrency return forecasting include sentiment analysis (Nasekin & Chen, 2020), semantic analysis (Kraaijeveld & De Smedt, 2020; Ortu, Uras, Conversano, Bartolucci, & Destefanis, 2022), and topic modeling (Loginova, Tsang, van Heijningen, Kerkhove, & Benoit, 2021).

We focus on sentiment analysis, a sub-field of NLP, which aims to identify the polarity of a piece of text (Pang, Lee, & others, 2008). Sentiment analysis methods have developed from lexicon-based approaches to state-of-the-art transformer models (Mishev, Gjorgjevikj, Vodenska, Chitkushev, & Trajanov, 2020). Given vast empirical evidence that sentiment scores extracted from investment-related social media and other online platforms reflect the subjective perception of investors, many studies investigate the potential value of adding corresponding features into (cryptocurrency) return forecasting models. Examining many cryptocurrencies, Nasekin and Chen (2020) show that sentiment from StockTwits extracted by BERT with added domain-specific tokens contributes to return predictability. Chen et al. (2019) use public sentiment from Reddit and StockTwits and confirm the value of sentiment features in a return prediction model, with the condition that the text classifier uses a domain-specific lexicon.

Polasik et al. (2015) find that Bitcoin prices are highly driven by news volume and news sentiment. Vo et al. (2019) show that sentiment scores extracted from news articles of the past seven days increase the predictive performance of an LSTM model in predicting cryptocurrency price directions. Ortu et al. (2022) conclude that features based on BERT-based emotion classification of GitHub and Reddit comments significantly improve the hourly and daily return direction predictability of Bitcoin and Ethereum.

A general challenge in forecasting using text data concerns the absence of labeled data. The training of a predictive model requires known outcomes of the target variable. Accordingly, developing a sentiment classifier requires text data with actual sentiment labels. Yet, news and social media data are naturally unlabeled. Prior work has considered various approaches to remedy the lack of labels or circumvent the challenge. These include manual labeling text by domain experts (Li, Bu, Li, & Wu, 2020; Malo, Sinha, Korhonen, Wallenius, & Takala, 2014; Van de Kauter, Breesch, & Hoste, 2015; Cortis, et al., 2017). Involving human labor, this approach lacks scalability. A more feasible, established alternative is to rely on sentiment dictionaries (Taboada, Brooke, Tofiloski, Voll, & Stede, 2011). However, the state-of-the-art in sentiment classification has moved far beyond lexicon-based approaches (Zhang, Wang, & Liu, 2018), which are criticized for being highly domain-dependent and lacking contextual understanding (Basiri & Kabiri, 2020). Therefore, the prevailing approach for extracting features from text data involves using pretrained embeddings from, e.g., Word2Vec (Rybinski, 2021) or BERT (Jiang, Lyu, Yuan, Wang, & Ding, 2022). Such embeddings are powerful but their understanding of language is based on the corpus on which they were (pre)trained. Without further finetuning, embeddings cannot accommodate peculiarities in a target corpus such as financial news or social media postings. To address the unlabeled data challenge, we introduce a recently proposed NLP approach called weak supervision to the forecasting literature. Weak supervision facilitates using unlabeled text data that represents the specific language in a target domain such as cryptocurrencies for finetuning pretrained embeddings and text classifiers like BERT. To our best knowledge, weak supervision has not been used in the field of financial - especially for cryptocurrency-related - forecasting.

# 3. Data

## 3.1. Text Data

This study focuses on sentiment analysis and financial prediction for Bitcoin and Ethereum, which are currently the two largest cryptocurrencies by market capitalization. Their popularity and relatively long period of existence motivate their choice for this study.

Bitcoin and Ethereum have their text datasets and are handled separately. Both datasets consist of news articles, Tweets, and Reddit posts in the period from 01/08/2019 until 15/02/2022 and scraped using different Python libraries, as shown in Table A1 in the online appendix.

The raw text datasets are filtered and prepared for analysis. The preparation differs across data sources, due to their unique format and varying ways of accessing them. News articles in the GoogleNews RSS feed are filtered to include only those published by CoinDesk and CoinTelegraph, and contain the name or code of the respective currency, i.e., "Bitcoin" or "BTC" and "Ethereum" or "ETH", adding up to, on average, 20 news articles per day and currency. There is no further filtration for the news data since there are no spam samples.

Twitter data is filtered by selecting and searching for the name and code of each currency from all Tweets that day. We then filter results based on a minimum threshold number of retweets, number of likes, number of followers of the corresponding account, and whether the account ID is verified. This helps to eliminate scams, advertisements, or irrelevant samples and emphasizes more publicly viewed posts published by relatively important accounts. The underlying assumption is that a post with a larger reach better reflects the community sentiment about cryptocurrency price changes.

For the Reddit data, the r/Bitcoin and r/ethereum subreddits are scraped and further filtered based on the number of likes and comments. Here, the assumption is that the priority lies in the commonly commented and discussed posts that likely have more critical content. Also, posts that are very short or just contain a URL are eliminated, as they would be meaningless to train or predict with the sentiment classification models.

There are common text samples between the Bitcoin and Ethereum datasets, especially in news and Tweets. The samples that contain both Bitcoin's and Ethereum's name or code are included in both datasets.

## 3.2. Financial Data

The target variable in the financial forecasting models is the daily rate of return. For each currency, the daily return at time t is computed as the percentage change from the closing price at time t to the closing price at time t+1. There is exactly one day between successive periods since data granularity is defined as daily for all features and the target. The beginning time of each day is taken as 00:00 Greenwich Mean Time, acting as the separation between the daily periods. Each period covers the entire day, until the time the next day starts.

Recent cryptocurrency price- or return- prediction models include the so-called price, technical, blockchain, macroeconomic, and other currency-related features (Jang & Lee, 2017; Huang, Huang, & Ni, 2019; Mallqui & Fernandes, 2019). Price features consist of Open-High-Low-Close (OHLC) data. Technical features express volume, volatility, trend, and momentum. Blockchain features include trading volume, block size, hash rate, and such blockchain-related information. Macroeconomic features consist of various indices such as S&P500 and external economic factors including inflation. Other currency features consist of price data of various cryptocurrencies, precious metals such as gold, or fiat currencies.

Table 1 reports the input features used in this paper, which we categorize into price, blockchain, macroeconomic, technical, sentiment, and dummy features. We also include weekday dummies to account for any weekly seasonality. All features in Table 1 are included for both Bitcoin and Ethereum return prediction models. The blockchain features are an exception as they are freely available for Bitcoin only. It was not possible to find similar and freely available network data for Ethereum. We use the Python wrapper for the open-source TA-Lib (Benediktsson & Cappello, 2022) to compute the technical features shown in Table 1. They depend only on the past prices of the corresponding currency. The sentiment features are computed using the predictions of the best-performing sentiment classifier selected in Section 4. How daily sentiment scores are then aggregated is explained in depth in Section 5.2 of the paper.

The one- and two-day lags of some features are also added to the models, as shown in Table 1. Given that we aim to predict the return of the following day, two lags appear sufficient. In total, including the lagged versions of features, there are 178 and 151 features for BTC and ETH models, respectively.

We do not use additional lags of the sentiment-related features. These features are inherently lagged since we aggregate the sentiment from the posts in the last 24 hours to form a daily score, as explained further in Section 5.2. The efficient market hypothesis indicates that predictable patterns in price developments – if they exist – will vanish quickly due to the immediate adaptation of prices when new information is revealed (Jordan, 1983). Likely, news and social media posts would already spread to a significant follower body in one day. Therefore, we assume that one day is already a long period for cryptocurrency prices to incorporate external effects, making it irrelevant to further lag sentiment-related features.

| Feature type | Number of features (without lags) | Feature list | Lags |
|---|---|---|---|
| Price features | 5 | Open, high, low, close, and adjusted close prices | 0 |
| Blockchain features* | 9 | Average block size 7-day MA, estimated transaction volume, hash rate 7-day MA, market capitalization, miners' total revenue, number of total transactions, number of total transactions 7-day MA, network difficulty, number of transactions per block | 0, 1, 2 |
| Macroeconomic features | 5 | 5-year breakeven inflation, 7-month treasury bill, S&P 500 close price, return of S&P 500 close price, volatility index of S&P 500 | 0, 1, 2 |
| Technical features (lagged) | 10 | Daily return between closing and opening prices, log of daily return, cumulative return, trading volume, price volatility by 30-day moving standard deviation of price, Parkinson volatility, relative intraday price change, closing price of the other cryptocurrency, daily return on the other cryptocurrency, volume of the other cryptocurrency** | 0, 1, 2 |
| Technical features*** (non-lagged) | 78 | **Volume:** money flow index, accumulation/distribution index, on-balance volume, Chaikin money flow, force index, ease of movement, volume-price trend, negative volume index, volume weighted average price<br>**Volatility:** average true range, Bollinger bands, Keltner channel, Donchian channel, Ulcer index<br>**Trend:** simple moving average, exponential moving average, weighted moving average, moving average convergence divergence (signal line, histogram), average directional movement index, vortex indicator (positive, negative, difference), Trix, mass index, commodity channel index, detrended price oscillator, KST oscillator, Ichimoku Kinko Hyo (span A, span B), parabolic stop and reverse (up, down), Schaff trend cycle<br>**Momentum:** relative strength index, stochastic relative strength index, true strength index, ultimate oscillator, stochastic oscillator, Williams %R, awesome oscillator, Kaufman's adaptive moving average, rate of change, percentage price oscillator (signal, histogram), percentage volume oscillator (signal, histogram) | 0 |
| Sentiment features | 6 | Daily count of news articles, count of Tweets, count of Reddit posts, daily overall sentiment score of news, score of Tweets, score of Reddit posts | 0 |
| Dummy features | 7 | Day dummies | 0 |

*Blockchain features are only included in the Bitcoin models*
*\*\* Other cryptocurrency refers to Bitcoin for the Ethereum models and vice versa*
*\*\*\* Computed using the open-source library TA-Lib (Benediktsson & Cappello, 2022)*

Table 1: List of all features in the financial prediction models

The data is split into training and test datasets, such that the training data is from 01/08/2019 until 31/07/2021, and the test data is from 01/08/2021 until 15/02/2022. The test period covers both a bullish and bearish pattern, to prevent the prediction and investment gain outcomes from being biased due to a single-direction price movement during the test period. Figure 1 shows the daily closing prices of Bitcoin and Ethereum during the training and test periods. The training and test datasets are standardized using the mean and standard deviation characteristics of the training data. Standardization is applied to all features with continuous values.

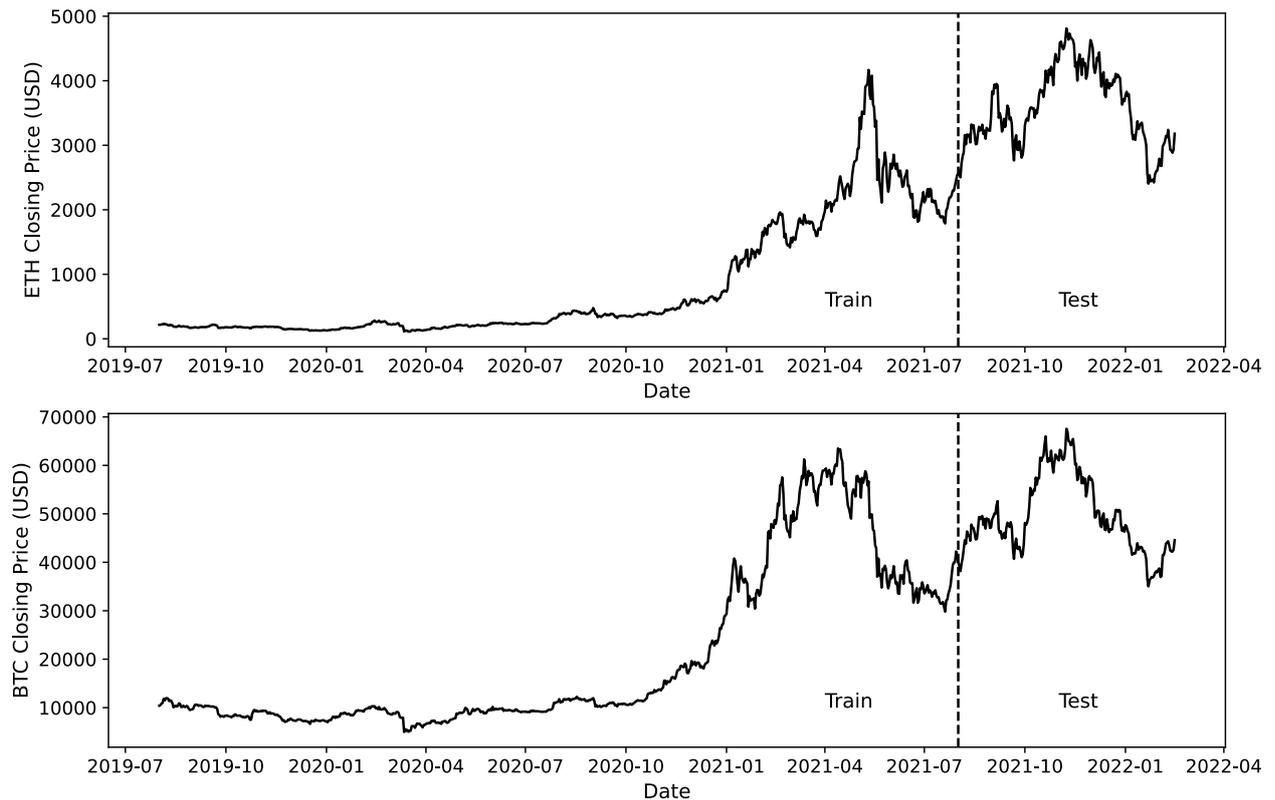

Figure 1: Training and test data split for Bitcoin and Ethereum price data

## 4. Sentiment Classification Models

### 4.1. BERT-Based Models

BERT (Devlin, Chang, Lee, & Toutanova, 2018) is an NLP model pretrained on a very large corpus. Shortly after its introduction in 2019, BERT quickly gained popularity due to its ease of use, coverage in several libraries, and excellent empirical results in many domains. Recent work on NLP applications in finance also uses BERT (Jiang, Lyu, Yuan, Wang, & Ding, 2022; Stevenson, Mues, & Bravo, 2021), which further motivates its choice for this study.

We build BERT-based models to classify the sentiment of news and social media data. The BERT classifiers consist of a pretrained uncased BERT base model with an additional classification layer. This final layer predicts sentiment as positive, neutral, or negative.

We finetune the BERT classifiers on the cryptocurrency-related text data to tailor the models to our specific context. Specifically, we build three variations of the sentiment classifiers:

- BERT-Frozen: Finetuned by freezing all pretrained BERT layers and only updating the weights in the final classification layer
- BERT-Unfrozen: Finetuning involves changing the weights in all the pretrained BERT layers and the final classification layer
- BERT-Context: Same as BERT-Unfrozen, but with additional cryptocurrency-related contextual tokens added to the tokenizer. The weights of the new tokens are initialized and updated during finetuning based on the specific contextual text data.

BERT-Frozen is the least elaborate sentiment classifier. Keeping the pretrained BERT layers unchanged limits the classifier's ability to learn from the target corpus. BERT-Unfrozen tackles this issue by updating BERT layers as well as the classification layer during training. This enables the BERT layers to adjust to the target corpus and deliver more relevant information to the final classification layer.

The vocabulary of the pretrained BERT model does not include cryptocurrency-related words such as *Bitcoin*, *Blockchain*, or even the word *cryptocurrency* itself. Therefore, the first two BERT-based classifiers cannot make sense of such contextual vocabulary as a whole word. Instead, they consider corresponding terms as unknown words and break them up into subwords like *bit* and *coin*, the semantics of which differ from the original token *bitcoin*. By adding contextual tokens to the BERT vocabulary, we enhance the ability of BERT-Context to learn patterns from the target corpus and extract cryptocurrency-related sentiment. Based on these considerations, we expect BERT-Frozen and BERT-Context to provide the least and most accurate sentiment scores, respectively.

### 4.2. Ensemble Sentiment Classifiers

Ensemble models can deliver better performance by synthesizing several models (Polikar, 2012, pp. 1–34; Lin, Kung, & Leu, 2022; Clemen, 1989). We combine the output of multiple selected text models to classify sentiment more accurately. The final prediction of an ensemble is computed by taking the majority vote of the single models. The selected ensembles are defined in upcoming subsections, where the model outputs are described.

There are some specifications when computing the majority vote of the models involved in an ensemble model. In the case that there is a draw between the number of negative and positive prediction votes, the ensemble predicts a neutral label. However, the ensemble output is less clear in the case of a draw between neutral and either positive or negative votes. We refer to an ensemble that predicts a neutral label in this case as "neutrality-biased" (NB) ensemble. Likewise, we call an ensemble that predicts either a positive or a negative label a "polarity-biased" (PB) ensemble.

### 4.3. Unlabeled Data and Weak Supervision

Finetuning a pretrained BERT model on our news and social media data is useful to adjust the sentiment classifier to the linguistic peculiarities that this data exhibits. However, finetuning is impossible without actual sentiment labels. We propose solving this problem by generating pseudo-labels from a zero-shot classifier without demanding any human supervision. Zero-shot classification (ZSC) involves a model that performs single- or multi-class labeling in an unseen domain with no prior training or predefined classes (Chang, Ratinov, Roth, & Srikumar, 2008; Xian, Lampert, Schiele, & Akata, 2018). We use a ZSC model to predict classes for our text data and then use those predictions as pseudo-labels to finetune our BERT-based classifiers. The ZSC model of choice is a pretrained BART multi-genre natural language inference (MNLI) model (Lewis, et al., 2019). The BART-based ZSC predicts the probability of a hypothesis defined for a piece being correct in an unsupervised manner (Yin, Hay, & Roth, 2019). In our case, there are three hypotheses for all samples, correspond to the classes positive, neutral, or negative sentiment. The hypothesis with the highest likelihood is the predicted class for the corresponding text sample. We use the ZSC predictions as pseudo-labels, or "weak labels", and employ them for finetuning BERT-based sentiment classifiers. This technique is called weak supervision (Zhou, 2018). Various studies confirm the merit of weakly-supervised learning in different NLP tasks and conclude that weak supervision outperforms the sole use of unsupervised models (Lison, Hubin, Barnes, & Touileb, 2020; Xiong, Du, Wang, & Stoyanov, 2019; Mekala & Shang, 2020; Popa & Rebedea, 2021). This motivates us to examine whether the encouraging results of weak supervision extend to cryptocurrency return prediction.

### 4.4. Weak Labeling Experiment

FinBERT (Araci, 2019) is a BERT-based sentiment classifier that was finetuned on the Financial Phrasebank data (Malo, Sinha, Korhonen, Wallenius, & Takala, 2014), a 4840-sample financial news dataset labeled by the majority vote from 16 individuals with adequate financial knowhow. The sentiment labels are either positive, neutral, or negative. We use this expert-labeled dataset as reference data to compare the reliability of weak labeling using pseudo-labels. To ensure comparability, we partition the Phrasebank data into train, validation, and test sets exactly as defined in the FinBERT finetuning algorithm (Araci, 2019).

Firstly, the entire dataset is classified using the BART ZSC model (as introduced in Section 4.3), generating labels with 79.0% accuracy compared to the expert labels. Next, we use the predicted labels as pseudo-labels to finetune BERT-Frozen and BERT-Unfrozen; just as if the data was unlabeled. For each of the two classifiers, we tune hyperparameters using grid search based on accuracy, precision and recall on the validation dataset. The best hyperparameters are used to train a final classifier on the union of the train and validation sets. Noting that our approach delivers multiple predictions per text sample and that these can be combined in different ways, we also introduce two ensemble classifiers (see Section 4.2 for details):

- Ensemble ZU-nb: Neutrality-biased majority vote from BART ZSC and BERT-Unfrozen
- Ensemble ZFU: Majority vote from BART ZSC, BERT-Frozen, and BERT-Unfrozen

| Model | Accuracy | Precision | Recall | F1 Score |
|---|---|---|---|---|
| BART ZSC | 0.790 | 0.775 | 0.771 | 0.773 |
| FinBERT | 0.822 | 0.787 | 0.823 | 0.805 |
| BERT-Frozen | 0.688 | 0.622 | 0.583 | 0.602 |
| BERT-Unfrozen | 0.789 | 0.769 | 0.754 | 0.761 |
| Ensemble ZU-nb | 0.795 | 0.816 | 0.717 | 0.763 |
| Ensemble ZFU | 0.787 | 0.784 | 0.737 | 0.760 |

Table 2: Sentiment classification performance metrics with respect to the actual labels

Table 2 shows the performance of all models based on the 970 test samples in terms of accuracy, unweighted macro-averaged precision, recall, and the F1 score. We chose the unweighted macro-averaging approach to avoid prioritizing neutral cases, which represent 59.3% of the test set, whereas the negative and positive class cover 13.2% and 27.5%, respectively.

BERT-Unfrozen, trained on the ZSC predictions as pseudo-labels, has a classification accuracy of 78.9%, which is not much lower than the 82.2% accuracy of FinBERT, which is trained on the actual labels. In terms of accuracy, Ensemble ZU-nb is the best model with 79.5% accuracy and performs the closest to FinBERT.

Table 3 shows the confusion matrices for the predictions of FinBERT and BERT-Unfrozen, compared to the actual labels. We note that BERT-Unfrozen successfully distinguishes between positive and negative samples, as does FinBERT. Most of the inaccuracy comes from incorrect labeling of negative or positive samples as neutral and vice versa.

|  | FinBERT Predictions | | |
|---|---|---|---|
|  | Pos. | Neg. | Neu. |
| Pos. | 22 | 0.72 | 4.5 |
| Neg. | 0.1 | 11 | 2.1 |
| Neu. | 6.8 | 3.6 | 49 |

(a)

|  | BERT-Unfrozen Pred. | | |
|---|---|---|---|
|  | Pos. | Neg. | Neu. |
| Pos. | 20 | 0.93 | 6.5 |
| Neg. | 0.41 | 9.2 | 3.6 |
| Neu. | 7.8 | 1.9 | 50 |

(b)

Table 3: Confusion matrices (a) for FinBERT and (b) for BERT-Unfrozen predictions, in percentages

So far, we have measured accuracy based on the test samples for which at least eight of the 16 expert labelers agreed on a single label. Table 4 presents the classifier on the 433 test samples that all of the 16 expert labelers agreed on. There, the accuracy of FinBERT and BERT-Unfrozen increases substantially to 92.6% and 91.7%, respectively. Ensemble ZFU even outperforms FinBERT based on the F1 score, indicating a better combined precision-recall performance, and achieves a very similar accuracy score of 92.4%.

| Model | Accuracy | Precision | Recall | F1 Score |
|---|---|---|---|---|
| BART ZSC | 0.919 | 0.896 | 0.925 | 0.910 |
| FinBERT | 0.926 | 0.900 | 0.902 | 0.901 |
| BERT-Frozen | 0.783 | 0.695 | 0.701 | 0.698 |
| BERT-Unfrozen | 0.917 | 0.889 | 0.890 | 0.889 |
| Ensemble ZU-nb | 0.917 | 0.930 | 0.871 | 0.900 |
| Ensemble ZFU | 0.924 | 0.916 | 0.896 | 0.906 |

Table 4: Performance metrics of the sentiment classifiers on the samples that all labelers agree

|  | FinBERT Predictions | | |
|---|---|---|---|
|  | Pos. | Neg. | Neu. |
| Pos. | 21 | 1.6 | 2.1 |
| Neg. | 0 | 12 | 1.4 |
| Neu. | 1.2 | 1.2 | 60 |

(a)

|  | Ensemble ZFU Pred. | | |
|---|---|---|---|
|  | Pos. | Neg. | Neu. |
| Pos. | 20 | 1.4 | 3.5 |
| Neg. | 0 | 12 | 1.2 |
| Neu. | 1.6 | 0 | 61 |

(b)

Table 5: Confusion matrices (a) for FinBERT and (b) for Ensemble ZFU predictions, in percentages

Table 5 portrays confusion matrices for the predictions of FinBERT and Ensemble ZFU on the selected samples. Regarding the significant performance increase of the models after sampling texts that all labelers agree on, the cases that the models misclassify are those on which even the human labelers disagree. This happens mostly between negative-neutral and positive-neutral labels. It is plausible that defining a clear boundary between these label pairs is a difficult task; even for a human. In sum, the experiment suggests that BERT-based sentiment classification models finetuned on weak labels of ZSC predictions can perform competitively to a model trained on the same data with the actual labels. We consider this an encouraging result and further evidence for the potential value of the weak-supervision approach.

### 4.5. BERT-Based Models for Cryptocurrency Data

Having confirmed the applicability of weak labeling unlabeled data in a financial context, we now generalize this outcome to our custom corpus related to cryptocurrencies. Specifically, we classify news articles, Reddit posts, and Tweets for Bitcoin and Ethereum between 01/08/2019 and 31/07/2021, corresponding to the training period for the financial prediction models, using BART ZSC and FinBERT. We then employ those labels to finetune our three BERT-based classifiers for sentiment prediction. To reduce the number of training samples, we filter the cases for which BART ZSC and FinBERT predict the same label.

We aim at finetuning BERT models for the task of classifying the sentiment from all text sources related to both cryptocurrencies. This suggests that the training dataset should give equal weight to each type of text. To achieve this, we further sample the data to obtain a 6,022-sample subset that is balanced in terms of sentiment class, source type, and related cryptocurrency.

We identify words that are not included in the default BERT vocabulary and order these by their frequency of occurrence. We then manually add relevant tokens from the most frequent words to BERT's token list. The added tokens are *btc*, *bitcoin*, *eth*, *ethereum*, *crypto*, *cryptocurrency*, *blockchain*, *defi*, *nft*, *binance*, *bullish* and *bearish*, all being in lowercase since we use uncased BERT. These tokens appear 50 to 1000 times in the training data, which might suffice to learn corresponding embeddings during finetuning.

BERT-Frozen, BERT-Unfrozen, and BERT-Context are finetuned on the weakly-labeled cryptocurrency text dataset. Firstly, hyperparameter tuning is performed using grid search and the best set of hyperparameters is selected according to a combined score of accuracy, precision, and recall metrics on the validation dataset. A final model is then trained on the entire training and validation dataset using the optimal hyperparameters. We repeat this process for each of the three BERT-based classifiers to obtain their final and optimized versions.

## 4.6. Model Performance Evaluation by Manual Labeling

The optimized BERT-based sentiment classifiers should be compared in terms of their classification performances. However, this is impossible as true sentiment labels are not available for our cryptocurrency corpus. To approximate classifier performance, a small but representative subset of the data is manually labeled by three individuals, consisting of an economics Ph.D. candidate, an economics master's graduate, and the first author of the paper. We select validation set samples for manual labeling according to the following protocol. First, we draw a balanced sample in terms of sentiment class (as estimated by BART), text source, and cryptocurrency; as previously done for the training data. Further, we adjust the sample such that a third of the selected cases consists of text pieces that all models agree on, another third consists of samples that BERT-Context differs from the weak labels it trained on, and the rest are cases for which the three BERT models predict different classes. With these steps, we aim to make the small manually labeled test sample as representative as possible of the whole dataset.

The human labelers agree on 78.0% of the samples. The real label of each sample is defined by the majority vote of the three raters. Table A2 in the online appendix offers some exemplary texts together with the manually assigned sentiment labels to clarify our approach.

| Model | Accuracy | Precision | Recall | F1 Score |
|---|---|---|---|---|
| BART ZSC | 0.587 | 0.647 | 0.621 | 0.634 |
| FinBERT | 0.580 | 0.579 | 0.575 | 0.577 |
| BERT-Frozen | 0.517 | 0.542 | 0.563 | 0.552 |
| BERT-Unfrozen | 0.580 | 0.580 | 0.622 | 0.600 |
| BERT-Context | 0.611 | 0.605 | 0.636 | 0.620 |
| Ensemble ZUCF-pb | 0.700 | 0.685 | 0.708 | 0.696 |

Table 6: Performance metrics of the models based on the manually classified labels

Table 6 provides the sentiment classifier comparison. First, we note that BERT-Context performs best among the three finetuned BERT models in accuracy and F1 score. Given that BERT-Context can process some cryptocurrency-specific vocabulary, whereas BERT-Frozen and BERT-Unfrozen can only use subword embeddings for cryptocurrency jargon, this result agrees with our expectations. More importantly, Table 6 reemphasizes the merit of weak supervision for finetuning BERT. BART ZSC and FinBERT perform sentiment classification with 58.7% and 58.0% accuracy. BERT-Context, which is trained on a dataset that incorporates the predictions of BART ZSC as pseudo-labels, achieves a higher accuracy of 61.1%. This evidences that compared to using an unsupervised model or a model trained on a different kind of data, it is beneficial to finetune text classifiers using weak labels.

Table 6 also includes the best-performing ensemble is included in the comparison, Ensemble ZUCF-pb, which is given by a polarity-biased majority vote from BART ZSC, BERT-Unfrozen, BERT-Context, and FinBERT. Ensemble ZUCF-pb outperforms all single models as well as other considered ensembles not included in Table 6 in terms of both accuracy and F1 score. The confusion matrix in Table 7 shows that this ensemble can distinguish positive and negative classes rather successfully, whereby most errors come from the neutral class.

|  |  | Ensemble ZUCF-pb | | |
|---|---|---|---|---|
|  |  | Pos. | Neg. | Neu. |
| Manual Labels | Pos. | 35 | 3.5 | 11 |
|  | Neg. | 1.3 | 15 | 3 |
|  | Neu. | 7.2 | 4.3 | 20 |

Table 7: Confusion matrix for Ensemble ZUCF-pb based on the manually classified labels, in percentages

We acknowledge that the above results are based on the specific, small, manually labeled subset and our specific sampling protocol. Therefore, the following analysis clarifies the merit of weak supervision in a financial forecasting context while using the entire test set. Given that Ensemble ZUCF-pb is the best sentiment classifier, we use its predictions to compute sentiment features for the financial prediction models, in the upcoming Section 5.

## 5. Financial Prediction Models

In this section, different regression and classification models are built to predict daily cryptocurrency returns. The purpose is to examine the contribution of public sentiment to the predictive performance of the forecasting models and potential gains from simulated trading.

### 5.1. Target and Features

The financial predictors are implemented separately for Bitcoin and Ethereum. The target and the features are explained in Section 3.2. Regression models predict the value of daily returns, whereas classification models classify returns as positive or negative.

### 5.2. Daily Sentiment Aggregation

Price movement prediction using investor sentiment poses a challenge due to the varying granularity and frequency of the sentiment features and the target. Unlike price or returns data, news and social media posts are distributed irregularly throughout time. Since we define our target as daily returns, we aggregate post-level sentiment scores into a daily average, which we then use as a feature in return prediction models. Previous research suggests alternative aggregation approaches. Some studies compute the logarithmic ratio of bullish to bearish posts per day (Antweiler & Frank, 2004), the ratio of the exponential moving averages of bullish to bearish cases to reduce noise (Leow, Nguyen, & Chua, 2021), or simply the ratio of the number of positive to negative posts (Bollen, Mao, & Zeng, 2011). In this paper, we take an aggregation approach that involves taking the difference between positive to negative samples and scaling it by the total number of posts that day, including neutral samples (Hiew, et al., 2019). Equation (1) details the calculation, whereby the post-level sentiment label predictions stem from the best-performing BERT-based Ensemble ZUCF-pb model in Section 4.

$$Sentiment\ Score_{t,s} = \frac{Pos_{t,s} - Neg_{t,s}}{Pos_{t,s} + Neu_{t,s} + Neg_{t,s}} \qquad (1)$$

Each positive, neutral or negative sample is considered to add 1, 0, or -1 points, respectively, to the daily score, which is then scaled by the number of text samples that day. We also include the daily text sample count as a feature in the financial model, as done by Bollen et al. (2011).

## 5.3. Evaluation of Sentiment Features Using VIF

The variance inflation factor (VIF) measures the extent to which a feature can be explained as a linear combination of the other features (Kennedy, 2008, p. 199). VIF values range from one, equivalent to no multicollinearity, to infinity, indicating perfect multicollinearity. It is commonly accepted that VIF between 5 and 10 indicates moderate collinearity and that there is severe multicollinearity if VIF is above 10 (Shrestha, 2020; Chatterjee & Hadi, 2006, pp. 233-239). We use VIF as a first test to examine whether sentiment-related features add unique information that other features do not contain. To that end, we perform feature elimination by computing VIF for each feature, removing the one with the highest value, and repeating until no feature has VIF above a cutoff. We consider cutoff values of 5 and 2.5. Table 9 shows the sentiment-related features that remain after iteratively removing all features with VIF values above 5. From the 178 and 151 features for Bitcoin and Ethereum, respectively, 46 and 42 features remain. Table 9 also marks the features that remain when the maximum VIF threshold is lowered to 2.5, leading to 34 and 32 features for Bitcoin and Ethereum, respectively.

|  | BTC | | | ETH | | |
|---|---|---|---|---|---|---|
|  | **News** | **Reddit** | **Tweets** | **News** | **Reddit** | **Tweets** |
| Count Features | No | Yes* | No | Yes | Yes* | Yes |
| Sentiment Features | Yes* | Yes* | Yes* | Yes* | Yes* | Yes* |

*\* Indicates features that are still not eliminated at a VIF cutoff of 2.5*
Table 8: Remaining features after elimination by VIF with a cutoff value of 5

For both cryptocurrencies, all sentiment features are retained after feature elimination, even at a VIF threshold of 2.5. This means that the news, Reddit, and Tweets sentiment features cannot be explained as linear combinations of the other features. Neither can they be explained by a linear combination of each other.

Out of all sentiment-related count features, Reddit post count is the only one that does not get eliminated at a VIF level of 2.5. This suggests that the Reddit count feature contributes to the feature set. The news and Tweets counts are, on the other hand, correlated with other features and, therefore, do not bring much additional linear value to the feature set.

### 5.4. Return Prediction Models

The regression and classification models shown in Table 10 are implemented separately for each cryptocurrency. The purpose is to evaluate the contribution of sentiment features to the predictability of future returns. A variety of models are built to minimize the dependency of the observed results on specific model choices and characteristics.

| Regression Models | Classification Models |
| --- | --- |
| - Ridge Regression | - Logistic Regression |
| - Support Vector Regressor | - Support Vector Classifier |
| - Multi-Layer Perceptron Regressor | - Linear Perceptron Classifier |
| - Stochastic Gradient Descent Regressor | - Multi-Layer Perceptron Classifier |
| - Decision Tree Regressor | - Decision Tree Classifier |
| - Random Forest Regressor | - Random Forest Classifier |
| - AdaBoost Regressor | - Extreme Gradient Boosting (XGBoost) Classifier |
| - Extreme Gradient Boosting (XGBoost) Regressor | - Light GBM Classifier |
| - Light GBM Regressor | - Long Short-Term Memory (LSTM) Classifier |
| - Voting Ensemble Regressor | - Gated Recurrent Unit (GRU) Classifier |
| - Linear Regression Ensemble Stacking Regressor | - Voting Ensemble Classifier |

Table 9: List of classification and regression models for daily return prediction

Each model is built two times, such that one is trained on the entire feature set and the other with all features except for the sentiment features. Fitting the 22 models of Table 10 for each of the two feature sets, which are then repeated for both cryptocurrencies, adds up to 88 models in total. Using these models, we compare models with and without sentiment features, as well as the respective performances of regression and classification models.

Ensemble models synthesize different machine learning models and are commonly built to raise forecast accuracy. We implement voting and stacking ensembles, which are included in Table 10. Although models such as XGBoost or random forest are also ensemble models, we refer only to the voting and stacking ensembles as *ensembles* in the rest of the paper. The ensemble regressors combine all regression models. Likewise, the ensemble classifier integrates all classifiers. The voting classifier predicts a class based on majority voting, whereas a voting regressor takes an unweighted average of the single model predictions. A stacking ensemble regressor combines single models by fitting a linear regression to their outputs.

We perform hyperparameter tuning separately for each Bitcoin and Ethereum model using stratified five-fold cross-validation with three repeats on the training data and select hyperparameters based on the balanced accuracy score. For regressors, this metric is computed by converting the predicted return values to binary classes. The selected hyperparameters with maximal cross-validated performance are used to fit final models on the entire training dataset. These models are tested on multiple test periods using an investment simulation.

## 5.5. Investment Simulation
### 5.5.1. Multiple Test Periods

The test period, throughout which the investment strategy runs, ranges from 01/08/2021 until 15/02/2022. During this timeframe, Bitcoin and Ethereum prices, in USD, change by 11.9% and 23.8%, respectively. The period contains a bullish pattern in approximately the first half and a bearish one in the other half. Figure 1 shows the closing prices in the test period.

Measuring the potential gains using a single test period would make the outcome susceptible to bias based on the specific price movements in the chosen period. To avoid this and ensure the robustness of results, we consider multiple shorter test frames within the entire test period to create a distribution of potential gains. The test period consists of 199 days. We take 60-day frames with starting dates shifted ten days forward each time, which leads to 14 different test periods. We compute the mean and standard deviation of the potential gains and compare models based on their mean performance.

It is important to note that we do not build a new model for each test frame. All test frames are forecasted using the same model built on the training set. Hence, the multiple test frame approach should not be confused with the sliding window approach often used in time series forecasting (Selvin, Vinayakumar, Gopalakrishnan, Menon, & Soman, 2017; Yang, Rivard, & Zmeureanu, 2005).

**5.5.2. Trading Strategy and Benchmark Scenarios**

Return prediction models can be difficult to interpret solely in terms of metrics such as accuracy. It is useful to also simulate the potential monetary gain. To that end, we conduct trading simulations separately for Bitcoin and Ethereum. We assume it is only possible to buy and sell one cryptocurrency using USD as fiat money. The trader's wallet contains 1000 USD and no cryptocurrency assets at the start of each test frame. Return, which is the target variable, is calculated as the percentage change in the closing price of the following day compared to that of the current day, at a 1-day horizon. Therefore, the trading strategy allows trading decisions to be made only at closing time every day, depending on the forecast direction of the closing price on the following day.

Our trading strategy is similar to that of Sebastião and Godinho (2021) and involves buying the corresponding cryptocurrency with all fiat money at the first time step at which a positive return for the next day is predicted. Then, the trader holds the assets as long as the price is predicted to rise the following day. On the first day that the closing price is predicted to fall, the trader sells the cryptocurrency and keeps the corresponding USD until a positive return is predicted for an upcoming day. At the end of the test period, the trader's wallet value is recorded in terms of USD. If the wallet contains cryptocurrencies, then their equivalent USD value is considered, for easier comparability to the input amount. The trading strategy aims to buy and sell at points where price movement directions change.

To set gains from the trading simulations into context, we define three benchmark cases representing an ideal, random, and holding scenario. The ideal scenario considers a trader who has complete foresight and always buys at peaks and sells at troughs (Lin, Yang, & Song, 2011; Jasic & Wood, 2004; Fernandez-Rodriguez, Gonzalez-Martel, & Sosvilla-Rivero, 2000). This way, one could capture all local minima and maxima as the price fluctuates, and consistently increase gains during the trading period. The transactions during the ideal scenario for investing in Bitcoin are displayed in Figure 2.

The random scenario considers a trader with no information about future market developments. We assign a random positive or negative return label to each day in the test periods and apply our trading strategy using these labels. This baseline is used to evaluate whether trading on model predictions delivers higher gains than random decision-making.

The passive buy-and-hold (B&H) strategy, where the trader buys at the beginning of a period and holds the asset until a predefined closing time, is also commonly used as a benchmark for price movement predictors (Leow, Nguyen, & Chua, 2021; Sebastião & Godinho, 2021). The percentage gain from this strategy is equivalent to the percentage price change of the corresponding cryptocurrency in terms of USD. This simulates a person who decides to invest in cryptocurrencies but does not make trading decisions based on opinions about short-term future price changes. The trader rather buys assets and waits, with the belief that this would bring positive returns over a longer horizon. We compare the trading gains from models to the B&H strategy to clarify whether capitalizing on short-term price fluctuations is feasible.

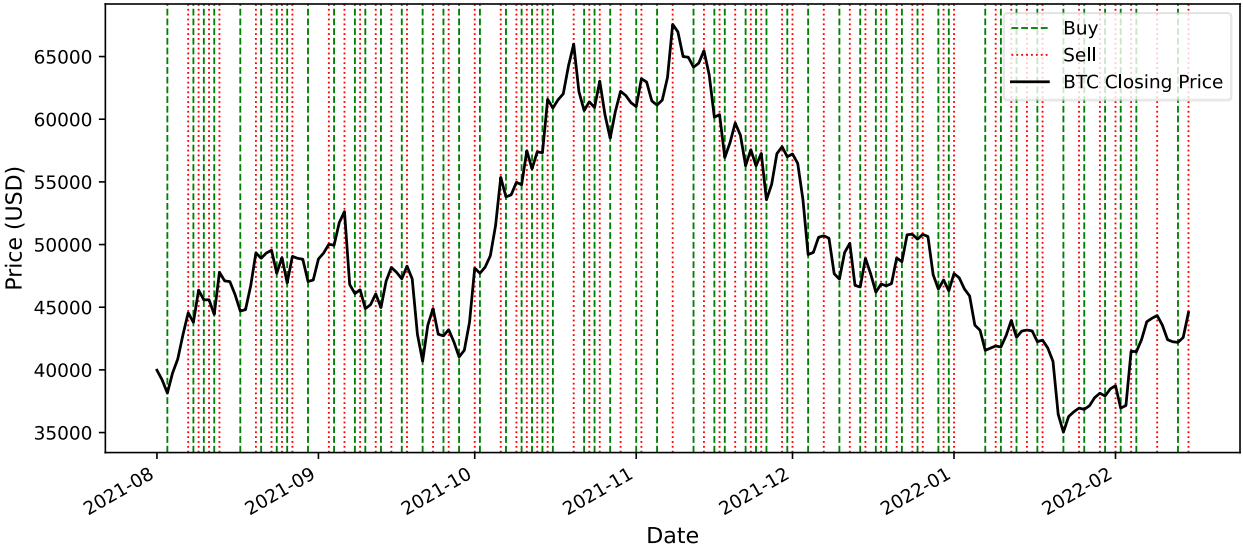

Figure 2: Ideal scenario in which a trader buys (sells) Bitcoin at local minima (maxima).

Most cryptocurrency trading platforms apply a percentage cost per transaction. Such transaction costs are important to consider in our trading simulation to make it closer to a real-life scenario. Transaction costs typically vary between 0.1% to 0.5% in the literature (Kim, 2017; Alessandretti, ElBahrawy, Aiello, & Baronchelli, 2018; Żbikowski, 2016, pp. 161-168). We consider a transaction cost of 0.2%.

The input amount is 1000 USD for each of the 14 test frames of 60-days. The output amount refers to the value of a trader's wallet in USD on the last day of each test period. Table 11 reports the result of the benchmarks. The values represent averages over the 14 test frames. Values in parentheses report the corresponding standard deviation.

|  | **For Bitcoin** | **For Ethereum** |
|---|---:|---:|
| Input Amount | 1000.00 (0) | 1000.00 (0) |
| Ideal Output Amount | 2011.80 (404.82) | 2494.01 (473.74) |
| Random Output Amount | 805.18 (216.63) | 900.23 (271.94) |
| Holding Output Amount | 1017.42 (291.39) | 1056.41 (292.37) |

Table 10: Ideal and baseline scenarios for Bitcoin and Ethereum. All values represent USD.

## 5.6. Model Performance Results and Discussion

### 5.6.1. Comparison of Best-Performing Models

The model performances and trading output for the best-performing ten models for Bitcoin and Ethereum, respectively, are displayed in Tables 12 and 13. These models provide the highest gain compared to the random baseline scenario. We detail the calculation of this relative gain below. Similar results for other models are available in the online appendix in Tables A3-A7.

The cross-validated training accuracy refers to the metric used during hyperparameter optimization; that is the five-fold stratified and three-times repeated cross-validated balanced accuracy score. For the test set, we report unbalanced accuracy because, in reality, one would not know the class distribution. Accuracy is directly calculated for classification models. For regression models, we transform the predictions into the corresponding positive or negative return. The accuracy measure is then calculated based on whether the regressor correctly predicts the direction of return.

From Tables 12 and 13, we observe that the mean number of transactions across the 60-day frames differs from 4 to 18. This equates to making a transaction every 3 to 18 days. The tables also report the trading costs, computed as 0.2% of each transaction.

We compare the results or trading model predictions to the benchmark scenarios (see Table 11). In Table 12 and Table 13, the column *Gains Scaled by Random Scenario* indicates the ratio of the gain from trading based on the model output compared to the gain from the random scenario. We compute this gain ratio for each of the 14 test periods separately. The computation leads to a distribution of the relative investment gain, consisting of these 14 observations. *Gains Scaled by Random Scenario* shows the mean and standard deviation, the latter being in parentheses, from this distribution. It is important to calculate the gain ratio separately for each test frame. Each frame exhibits a different price trend and, therefore, the distribution shows the overall performance as tested on a variety of possible future price changes. Similarly, *Gains Scaled by Hold Scenario* refers to the distribution of the gain from trading using model predictions as a ratio to the gain from the B&H strategy.

We analyze the degree of differentiation of our investment gains from the baseline scenarios. The t-values of the gain ratio distributions are computed to check for significance, with the null hypothesis being that the gain ratio is zero and the model does not add significant value compared to the baseline scenarios. The columns *t-value Random Scenario* and *t-value Hold Scenario* indicate the t-values of the distributions from *Gains Scaled by Random Scenario* and *Gains Scaled by Hold Scenario*, respectively. We use a two-tailed t-test to evaluate both directions of change from the baseline.

The highest test accuracy scores attained are 61.3% and 58.8% for Bitcoin and Ethereum models, respectively. Fama (1970) defines efficient capital markets as those in which the prices fully reflect the available information. With the assumption that financial markets should immediately respond to external factors under efficient conditions, predicting future asset prices should not be possible (Timmermann & Granger, 2004). This assumption, if true for the Bitcoin and Ethereum markets, would not allow for forecasts to be significantly better than a random guess, represented by the random baseline scenario. The best models for Bitcoin and Ethereum lead to gains of 0.572 and 0.501 in comparison to the random case, respectively. As displayed by the *t-value Random Scenario* values, the best ten models lead to significant improvements at the 90% confidence level, and most at the 99% confidence level. This result indicates that cryptocurrency markets are not fully efficient and there is room for successful predictions of future prices, which our predictive models achieve at a significant rate.

All models listed in Tables 12 and 13 deliver a mean result greater than 1000 USD, which is the input amount at the beginning of each test period. This implies that basing trading decisions on the model predictions leads to higher gains compared to not entering the cryptocurrency market at all. It is also observed that even the best-performing models for Bitcoin and Ethereum do not even nearly lead to gains as high as in the ideal scenarios shown in Table 11.

Trading based on the applied ten best-performing predictive models during the test periods consistently yields higher returns than the B&H strategy. However, only the best model for Bitcoin and the best two models for Ethereum outperform the holding scenario at a confidence level of 90%. Most of the models do not perform significantly better than B&H.

We find that models with the highest test accuracy do not necessarily yield the highest profit. While the support vector regressor has a higher test accuracy than ridge regression for Bitcoin, for example, the latter outperforms the former in terms of relative gains compared to the random scenario. This difference is caused by the specific trading strategy selected in this paper. A different trading strategy could have created a different ranking. The strategy we use values correct predictions at points of price direction change. Therefore, a model with a lower test accuracy can outperform another model with more accurate predictions, simply by correctly capturing periods that indicate a price direction change. This would lead to successfully buying at lower prices and selling at higher prices.

| Model Type | Model | Feature Set | Train CV Acc. | Test Acc. | Output Amount* (USD) | Gain Scaled by Hold Scenario* | t-value Hold Scenario | Gain Scaled by Random Scenario* | t-value Random Scenario | Total Trading Cost* (USD) | Num Tran.* |
|---|---|---|---|---|---|---|---|---|---|---|---|
| reg | ridge | all | 0.661 | 0.583 | 1204.90 (133.80) | 0.254 (0.255) | 0.049 | 0.572 (0.291) | 0.000011 | 29.47 (13.43) | 13 (5) |
| reg | mlp | no sent. | 0.639 | 0.583 | 1181.38 (119.45) | 0.235 (0.27) | 0.078 | 0.544 (0.296) | 0.000022 | 32.36 (6.06) | 14 (2) |
| reg | mlp | all | 0.648 | 0.573 | 1165.31 (120.65) | 0.226 (0.306) | 0.11 | 0.54 (0.361) | 0.000038 | 33.35 (8.13) | 15 (3) |
| reg | svm | all | 0.678 | 0.613 | 1174.86 (155.87) | 0.213 (0.215) | 0.1 | 0.518 (0.227) | 0.000044 | 40.23 (7.29) | 18 (2) |
| reg | ridge | no sent. | 0.641 | 0.558 | 1183.63 (202.02) | 0.211 (0.186) | 0.1 | 0.518 (0.202) | 0.000096 | 29.78 (11.07) | 13 (4) |
| reg | ens. vote | no sent. | 0.728 | 0.568 | 1124.43 (109.37) | 0.192 (0.326) | 0.23 | 0.497 (0.39) | 0.00014 | 33.84 (14.63) | 16 (6) |
| clf | lr | no sent. | 0.703 | 0.553 | 1163.76 (293.58) | 0.158 (0.07) | 0.21 | 0.459 (0.127) | 0.0017 | 25.56 (8.54) | 10 (3) |
| reg | ens. stack | all | 0.444 | 0.528 | 1085.10 (89.97) | 0.142 (0.274) | 0.44 | 0.424 (0.292) | 0.00046 | 8.39 (8.25) | 4 (4) |
| clf | per | no sent. | 0.632 | 0.568 | 1053.40 (69.88) | 0.121 (0.312) | 0.67 | 0.396 (0.338) | 0.0012 | 27.61 (9.30) | 13 (4) |
| reg | svm | no sent. | 0.636 | 0.553 | 1050.86 (93.05) | 0.122 (0.327) | 0.7 | 0.395 (0.353) | 0.0015 | 25.32 (8.92) | 12 (3) |

\* *Mean and standard deviation, the latter inside parentheses, are used to represent columns that refer to a distribution.*

Table 11: **BITCOIN** - Performance metrics and trading output of the ten best-performing models, in descending order

| Model Type | Model | Feature Set | Train CV Acc. | Test Acc. | Output Amount* (USD) | Gain Scaled by Hold Scenario* | t-value Hold Scenario | Gain Scaled by Random Scenario* | t-value Random Scenario | Total Trading Cost* (USD) | Num Tran.* |
|---|---|---|---|---|---|---|---|---|---|---|---|
| clf | mlp | all | 0.689 | 0.588 | 1293.98 (263.42) | 0.267 (0.202) | 0.039 | 0.501 (0.244) | 0.0009 | 27.52 (13.93) | 11 (5) |
| clf | svm | no sent. | 0.717 | 0.573 | 1285.68 (273.32) | 0.266 (0.242) | 0.049 | 0.494 (0.271) | 0.0013 | 35.43 (18.94) | 14 (7) |
| clf | ens. stack | all | 0.245 | 0.573 | 1243.33 (253.93) | 0.211 (0.147) | 0.094 | 0.447 (0.259) | 0.0026 | 39.76 (9.77) | 16 (3) |
| reg | mlp | all | 0.606 | 0.588 | 1181.46 (155.42) | 0.191 (0.285) | 0.19 | 0.416 (0.367) | 0.004 | 24.91 (14.25) | 11 (5) |
| reg | ridge | no sent. | 0.627 | 0.568 | 1188.31 (196.79) | 0.192 (0.284) | 0.19 | 0.412 (0.341) | 0.005 | 10.31 (10.00) | 4 (4) |
| clf | mlp | no sent. | 0.688 | 0.553 | 1177.64 (169.20) | 0.173 (0.238) | 0.21 | 0.396 (0.31) | 0.005 | 25.53 (11.74) | 11 (5) |
| reg | ridge | all | 0.630 | 0.573 | 1165.61 (188.26) | 0.169 (0.28) | 0.27 | 0.386 (0.342) | 0.0082 | 13.17 (6.94) | 6 (3) |
| reg | svm | all | 0.649 | 0.548 | 1155.45 (166.78) | 0.163 (0.284) | 0.3 | 0.38 (0.35) | 0.0087 | 8.26 (10.49) | 4 (4) |
| reg | ens. stack | no sent. | 0.546 | 0.568 | 1137.06 (143.07) | 0.15 (0.289) | 0.38 | 0.367 (0.371) | 0.012 | 19.63 (18.05) | 8 (7) |
| reg | tree | all | 0.736 | 0.518 | 1095.41 (73.55) | 0.134 (0.377) | 0.65 | 0.359 (0.489) | 0.025 | 13.00 (6.69) | 6 (3) |

\* *Mean and standard deviation, the latter inside parentheses, are used to represent columns that refer to a distribution.*

Table 12: **ETHEREUM** - Performance metrics and trading output of the ten best-performing models, in descending order

Out of the ten best-performing models for Bitcoin and Ethereum, four and six models, respectively, incorporate sentiment features. Also, the highest test accuracy is achieved by models that include sentiment features for both cryptocurrencies. These results indicate that the sentiment features are contributing to the predictive performance of the models, and more importantly to trading returns above the random baseline. Given that the features are generated from aggregated sentiment extracted from information with a relative long-term lag to the actual closing price used, these results are promising.

**5.6.2. Collective Evaluation of the Models**

Table 14 summarizes the output from all models to clarify the informational value of the sentiment features. The percentage of models in each category that outperform the random and holding baseline scenarios are shown in the columns *Outperf. Random Scenario* and *Outperf. Hold Scenario*, respectively. For example, 100% of all model types outperform the random scenario, meaning that all model predictions lead to a higher mean profit compared to the random case. The share of models that perform significantly better than baselines at a confidence level of 90% is also shown in the columns *Signif. Random Scenario* and *Signif. Hold Scenario*.

| Curr. | Model Type | Feature Set | Mean Train Acc. | Mean Test Acc. | Mean Output Amount | Outperf. Hold Scenario | Signif. Hold Scenario | Outperf. Random Scenario | Signif. Random Scenario |
|---|---|---|---|---|---|---|---|---|---|
| BTC | clf | all | 74.0% | 50.2% | **990.40** | **62.5%** | 0.0% | 100.0% | **81.2%** |
| | | no sent. | 76.0% | **51.4%** | 961.62 | 18.8% | **6.2%** | 100.0% | 37.5% |
| | reg | all | 69.6% | **52.9%** | 1057.91 | **84.6%** | **7.7%** | 100.0% | 92.3% |
| | | no sent. | 69.6% | 52.3% | 1033.42 | 76.9% | 0.0% | 100.0% | 92.3% |
| ETH | clf | all | 78.0% | 51.3% | 1057.01 | **62.5%** | 6.2% | 100.0% | **25.0%** |
| | | no sent. | 77.1% | **52.6%** | **1060.39** | 56.2% | 6.2% | 100.0% | 12.5% |
| | reg | all | 64.3% | **53.0%** | **1073.28** | 53.8% | 0.0% | 100.0% | **46.2%** |
| | | no sent. | 66.2% | 52.4% | 1039.79 | **84.6%** | 0.0% | 100.0% | 23.1% |

*Boldface indicates the better performing model per category (with/without sentiment)*
Table 13: Summary table comparing all models with and without sentiment features

Table 14 reveals that regressors tend to outperform classifiers for both cryptocurrencies. They tend to have higher mean test accuracy scores and are more likely to outperform the B&H baseline. Differences in the target variables may explain this result. Regressors train on the magnitude and direction of next-day returns, which facilitates differentiating between cases with small and large magnitudes of positive or negative returns. Classifiers, on the other hand, train on a binary target, which could result in information loss and inferior performance. Another striking difference between regression and classification results is that regressors show higher accuracy when sentiment features are included. Classifiers do not show this pattern. Instead, accuracy is slightly higher when sentiment features are excluded. When considering trading performance, however, we observe a higher share of models that incorporate sentiment features to outperform the random benchmark significantly. This means that, on average, sentiment features can add predictive information to the feature set. This relation does not hold for each of the models but it is valid when analyzing all models collectively.

Bitcoin models with sentiment features are more likely to outperform the holding scenario compared to those without. The same holds for Ethereum classifiers but not for Ethereum regressors, which outperform the B&H benchmark at a higher rate without sentiment. Only a negligible fraction of all models outperform the hold scenario significantly at a 90% confidence level. This indicates that profit differences between model-based trading and the B&H baseline are typically not substantial enough to conclude that the respective profit distributions differ. Using model predictions in decision-making supports the trader to get less affected by the rapidly fluctuating prices of cryptocurrencies and still adds some additional value compared to simply holding assets for a long term. For traders planning to invest in large volumes, every slight reduction in risk by having more accurate price predictions poses a profit opportunity.

By testing the value of sentiment on various models, we show that sentiment features contribute to higher investment gains, on average. It is important to conclude that this does not equate to a definite outcome that including news and social media sentiment-related features consistently leads to higher profits. Tables A3-A6 in the online appendix show that this does not hold for all models and cryptocurrencies. However, Table 14 summarizes that the models with sentiment features improve decision-making in cryptocurrency trading in a generalized manner. This suggests that public sentiment extracted from various online sources improves the predictive power of machine learning models to estimate future price direction changes of Bitcoin and Ethereum.

## 6. Conclusion

A key contribution of the paper is related to weak supervision. Textual data has become a common ingredient in (financial) forecasting models. While early approaches used dictionaries to extract features from text data, the use of pretrained embeddings has meanwhile become a de facto standard. In NLP, it is common practice to finetune a pretrained model on a corpus from the target domain. Finetuning accounts for linguistic peculiarities in that domain and, more generally, the fact that pretraining was carried out using text from an entirely different field that may have nothing in common with the target domain. An adaptation of a pretrained model to its target task (or dataset) is intuitively useful, but proved challenging in many forecasting settings, because most text data comes without labels. Weak supervision addresses this challenge. Forecasters can gather text data they judge important, use a cutting-edge pretrained NLP model, and finetune it before extracting features for their forecasting model. Using a unique multi-source dataset related to cryptocurrencies, we demonstrate the effectiveness of this approach in a sentiment classification context.

Beyond weak supervision and its empirical assessment, the paper also contributes to the empirical literature on cryptocurrency forecasting. Aggregating the sentiment classification from finetuned BERT using weak supervision, we investigate the added value of using investor sentiment for cryptocurrency return forecasting. The overall outcome is that sentiment features improve forecast accuracy across many tested forecasting models. Notably, we find strong evidence that several models, typically those that incorporate sentiment features, forecast cryptocurrency returns better than random and yield a positive trading profit after transaction costs. This suggests that Bitcoin and Ethereum markets cannot be considered efficient; at least for the period under study.